\documentclass[aps,floatfix,twocolumn,prb,superscriptaddress]{revtex4-2}
\usepackage{xcolor} 
\usepackage{graphicx}
\usepackage[hidelinks]{hyperref}
\usepackage{amssymb}
\usepackage[utf8]{inputenc}
\usepackage{amsmath}
\usepackage{braket}
\usepackage{subcaption}
\newcommand{\abs}[1]{\left|#1 \right|}

\renewcommand{\vec}[1]{\boldsymbol{#1}}

\definecolor{vastkust}{RGB}{0, 48, 80} 
\hypersetup{
  colorlinks,
  citecolor=vastkust,
  linkcolor=vastkust,
  urlcolor=vastkust}

\begin{document}

\title{Phonon Inverse Faraday effect from electron-phonon coupling}
\author{Natalia\ Shabala}
\email{natalia.shabala@chalmers.se}
\author{R.~Matthias\ Geilhufe}
\email{matthias.geilhufe@chalmers.se}
\affiliation{Department of Physics, Chalmers University of Technology, 412 96 G\"{o}teborg, Sweden}
\date{\today}

\begin{abstract}
The phonon inverse Faraday effect describes the emergence of a DC magnetization due to circularly polarized phonons. In this work we present a microscopic formalism for the phonon inverse Faraday effect. The formalism is based on time-dependent second order perturbation theory and electron phonon coupling. While our final equation is general and material independent, we provide estimates for the effective magnetic field expected for the ferroelectric soft mode in the oxide perovskite SrTiO$_3$. Our estimates are consistent with recent experiments showing a huge magnetization after a coherent excitation of circularly polarized phonons with THz laser light. Hence, the theoretical approach presented here is promising for shedding light into the microscopic mechanism of angular momentum transfer between ionic and electronic angular momentum, which is expected to play a central role in the phononic manipulation of magnetism. 
\end{abstract}

\maketitle

\section{Introduction}

Circularly polarized phonons or axial phonons are lattice vibrations with a non-zero angular momentum. These lattice vibrations can induce a magnetization in the material. This magnetization is an example of dynamical multiferroicity, the phenomenon in which the motion of ions in a crystal causes its polarization to vary in time, thus inducing a net magnetization \cite{rebane1983faraday,juraschek2017dynamical,juraschek2019orbital,geilhufe2021dynamically}. 

While the gyromagnetic ratio of the phonon hints towards a magnetization in the order of the nuclear magneton, recent experiments using phonon Zeeman effect and magneto-optical Kerr effect \cite{Basini2024,Luo2023, cheng2020large, baydin2022magnetic, hernandez2022chiral,Schaack1977} show that 
the size of the magnetization resulting from circularly polarized moment is quite significant, with magnetic moments on the order of magnitude of $0.1 - 10~\mu_\text{B}$. This is a promising route for using phonons for magnetic manipulation~\cite{Juraschek2020}, as has recently been shown on the example of the magnetic switching due to the ultrafast Barnett effect~\cite{Davies2024}. Here, the Barnett effect~\cite{Barnett1915} describes the magnetization of a nominally nonmagnetic sample due to mechanical rotation, and is the inverse of the so-called Einstein-de-Haas effect~\cite{Einstein1915}. Recently, both effects have been brought into the characteristic time and length-scales of material excitations, and phenomenologically describe the angular momentum transfer between magnetization and phonons~\cite{Zhang2014,Tauchert2022,Davies2024}.

These findings call for a microscopic theory of angular momentum transfer between phonons and electrons for describing the phonon-induced magnetic moments. As a result, multiple microscopic theories of this have been proposed, explaining the size of the phonon-induced magnetic moment, e.g., by inertial effects \cite{geilhufe2022dynamic,geilhufe2023electron}, orbit-lattice coupling \cite{chaudhary2023giant}, spin-orbit coupling \cite{fransson2023chiral}, orbital magnetization \cite{ren2021phonon}, electron-nuclear quantum geometry~\cite{Klebl2024}, non-Maxwellian fields~\cite{Merlin2023,Merlin2024}. While these approaches show similarities and overlap in their formalism, the consensus on the microscopic theory behind the effect has not yet been reached. 

In contrast, the optical analogue, i.e., the transfer of spin angular momentum from circularly polarized light to electron spin is well-described by the inverse Faraday effect~\cite{pershan1966theoretical}. Here, the electric field of the light couples to the electron via the dipole-interaction. From a symmetry perspective, the concept of inverse Faraday effect is universal and can be generalized to any circularly polarized vector field, beyond a laser field. Examples comprise axial magnetoelectric effect~\cite{liang2021} and the \textit{phonon inverse Faraday effect}~\cite{Juraschek2020}. Here, we provide the microscopic theory for the phonon inverse Faraday effect by coupling circularly polarized phonons to electrons via the electron-phonon interaction.

\begin{figure}
     \begin{subfigure}[b]{0.18\textwidth}
         \centering
         \includegraphics[width=\textwidth]{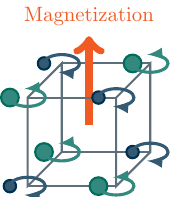}
         \caption{}
         \label{fig:ife}
     \end{subfigure}
     \hfill
     \begin{subfigure}[b]{0.26\textwidth}
         \centering
         \includegraphics[width=\textwidth]{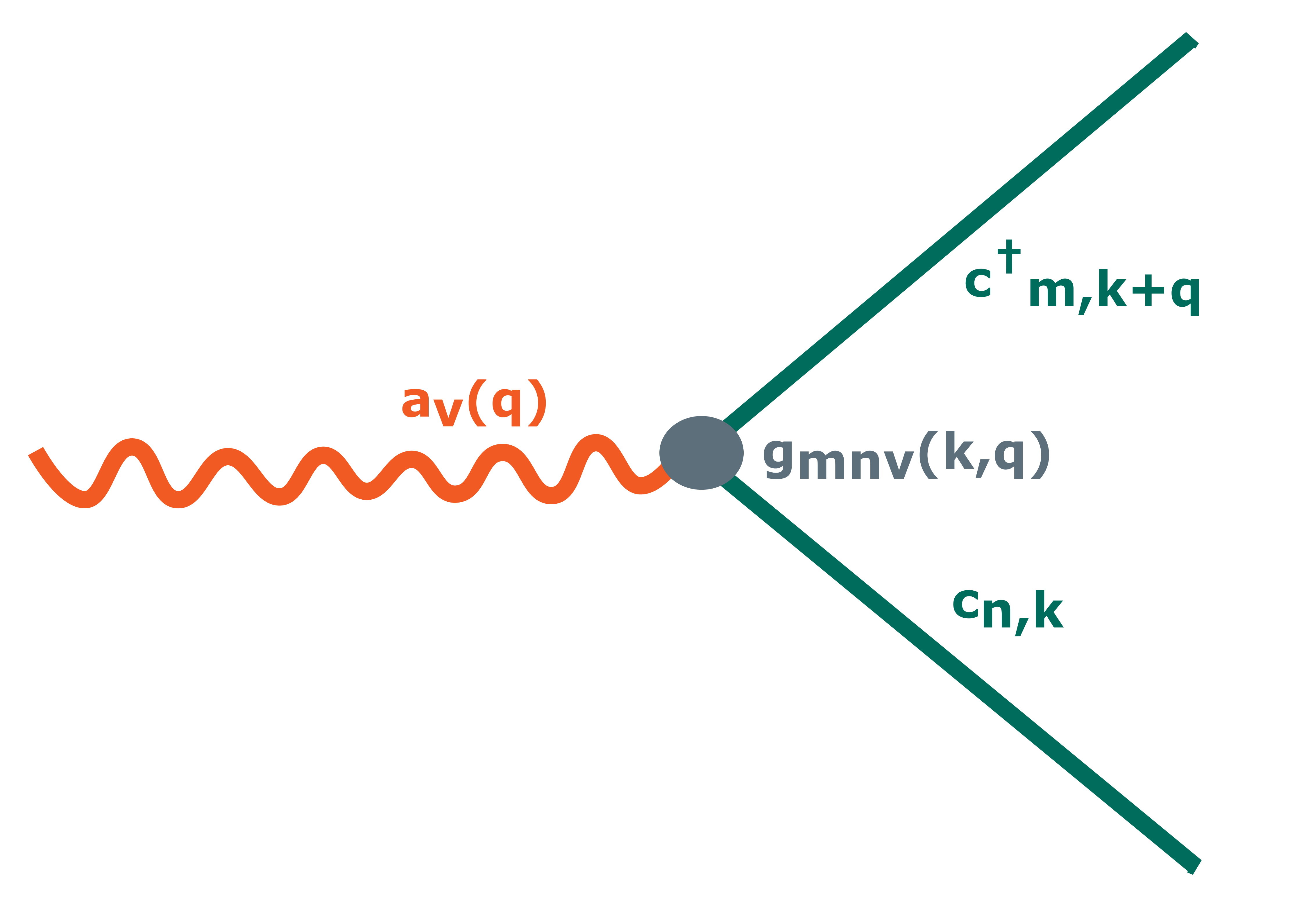}
         \caption{}
         \label{fig:vortex}
     \end{subfigure}
        \caption{a) Schematically depicted phonon inverse Faraday effect. b) Vortex diagram of electron-phonon interaction: electron in a state $\ket{n,\vec{k}}$ absorbs a phonon of mode $\nu$ with a wave vector $\vec{q}$ and scatters to a state $\ket{m, \vec{k}+\vec{q}}$}.
        \label{fig:ife_and_vortex}
\end{figure}

\section{Phonon inverse Faraday effect - Phenomenological theory}

We start with a phenomenological description of the phonon inverse Faraday effect, similar to Pershan \textit{et al.}
~\cite{pershan1966theoretical} and the optical inverse Faraday effect. For simplicity, we consider 2-fold degenerate phonon level, with two modes $u_\mu$ and $u_\nu$. We are free to introduce a basis transform, e.g., to the circularly polarized basis,
\begin{equation}
    \vec{u}(t) = \Big(\frac{1}{\sqrt{2}}u_R(\vec{\hat{e}}_\mu + i\vec{\hat{e}}_\nu) + \frac{1}{\sqrt{2}} u_L(\vec{\hat{e}}_\mu - i\vec{\hat{e}}_\nu)\Big)e^{i\omega t},
    \label{circularphonon}
\end{equation}
with $u_R = \left(u_\mu - i u_\nu\right)/\sqrt{2}$ and $u_L = \left(u_\mu + i u_\nu\right)/\sqrt{2}$. 
We need define free energy function in terms of phonon mode amplitudes. To fulfill the symmetry criteria of a nonmagnetic and inversion symmetric crystal, the thermodynamic free energy has to be invariant under time reversal and space inversion. This gives rise to the following phenomenological coupling betweeen circularly polarized phonons and the magnetic field $\vec{H}$,
\begin{equation}
    F_{u} = \chi H_z (u_R u_R^*-u_L u_L^*) = i \chi H_z (u_\mu u_\nu^*-u_\nu u_\mu^*).
\end{equation}
As a result, the magnetization is given by
\begin{equation}
        M_z = -\frac{\partial F_u}{\partial H_z} = \chi (u_L u_L^{*} - u_R u_R^*).\label{magnetization}
\end{equation}
From the equation above it becomes evident that an imbalance of circularly polarized phonons induce the DC magnetization of the material. Such an imbalance can be induced by coherent excitation with circularly polarized laser light~\cite{Luo2023,Basini2024,Davies2024}. However, the effect itself is purely phononic and does not require light. To offer a full picture, we give the phononic Faraday rotation,
\begin{align}
    \Delta\epsilon^u_R &= -4\pi \frac{\partial^2F_u}{\partial u_R \partial u_R^*} = -4\pi \chi H_z, \\     \Delta\epsilon^u_L &= -4\pi \frac{\partial^2F_u}{\partial u_L \partial u_L^*}= 4\pi \chi H_z.
    \label{eq:phononepsilon}
\end{align}
Hence, in the presence of an applied magnetic field, phonons develop circular polarization.

\section{Phonon inverse Faraday effect - microscopical theory}

In the following we develop the microscopic theory of the phonon inverse Faraday effect. A phonon is the collective excitation of the lattice, i.e., time-dependent displacements of the ions around their equilibrium positions. Hence, phonons introduce a time-dependent perturbation $V(t)$ into the system, $\hat{H} = \hat{H}_0 + V(t)$. We assume that the atom displacements $u_{pj\alpha}$ are sufficiently small and the potential function can be written as a first-order Taylor expansion. Then the perturbation $V(t)$ is given by
\begin{equation}
    V(t) = \sum_{pj\alpha}\frac{\partial U}{\partial u_{p j \alpha}}u_{pj\alpha},
\end{equation}
where we consider the variation of the potential due to displacement of atom $j$ in Cartesian direction $\alpha$ in a unit cell $p$. 

Before presenting the main result, we outline our approach for a single ion with displacement $\vec{u}(t)$. To allow for circular polarization, $\vec{u}(t)$ is generally complex. In this case the real-valued perturbation becomes $V(t)=2 \Re\left[\vec{u}(t)\cdot \nabla_{\vec{u}} U\right]$. Hence, we express the time-dependent perturbation as follows,
\begin{equation}
    V(t) = v e^{i \omega t} + v^*e^{-i\omega t},
    \label{perturbation}
\end{equation}
where $\omega$ denotes the phonon frequency. Equation \eqref{perturbation} gives rise to an effective Hamiltonian in second order perturbation theory~\cite{pershan1966theoretical},
\begin{multline} 
    \bra{a}H_{\text{eff}}(t)\ket{b} = -\sum_n \Big[\frac{\bra{a}v\ket{n}\bra{n}v^*\ket{b}}{E_{nb}-\hbar\omega} \\ - \frac{\bra{a}v^*\ket{n}\bra{n}v\ket{b}}{E_{nb}+\hbar\omega}\Big].\label{Heff}
\end{multline}
Here, $\ket{n}$ are eigenstates of the unperturbed Hamiltonian, and $E_{nb} = E_n - E_b$ denotes the energy difference between states $\ket{n}$ and $\ket{b}$.
We evaluate the effective Hamiltonian \eqref{Heff} and only keep terms giving a contribution to the magnetization as discussed in \eqref{magnetization}. This allows us to formulate the following revised effective Hamiltonian,
\begin{multline}\label{eq:H_eff_single_ion}
    \mathcal{H}_{\text{eff}}^{ab}(\vec{k}) =  -\hbar\omega (\vec{u} \times \vec{u^{*}})_z \times \\\times \sum_n \frac{(\bra{a}\nabla U\ket{n} \times \bra{n}\nabla U \ket{b} )_z}{E_{\vec{k}nb}^2-\hbar^2\omega^2}.
\end{multline}
Equation \eqref{eq:H_eff_single_ion} represents a semi-classical solution where the ionic displacement is not quantized. We generalize the single ion case for the entire crystal by introducing quantized normal coordinates for phonon modes $\mu$ and $\nu$: 
\begin{equation}\label{displacement_mu_nu}
\begin{split}
    & u_{p \mu} =  i\sum_{\vec{q}}e^{i\vec{q}\cdot \vec{R_p}}l_{\vec{q}\mu}(\hat{a}_{\vec{q}\mu}+\hat{a}^{\dagger}_{-\vec{q}\mu}), \\
    & u_{p \nu} =  \sum_{\vec{q}}e^{i\vec{q}\cdot \vec{R_p}}l_{\vec{q}\nu}(\hat{a}_{\vec{q}\nu}+\hat{a}^{\dagger}_{-\vec{q}\nu}).
\end{split}
\end{equation}
Here, $l_{\vec{q}\nu}=\sqrt{\frac{\hbar}{2\omega_{\vec{q}\nu}}}$ is the zero displacement amplitude. Operators $\hat{a}_{\vec{q}\nu}^{\dagger}$ and $\hat{a}_{\vec{q}\nu}$ are bosonic creation and annihilation operators. Together the phonon modes \eqref{displacement_mu_nu} form a circularly polarized phonon mode according to equation \eqref{circularphonon}. 

To describe the electron-phonon coupling we introduce electron-phonon matrix elements $g_{mn\nu}(\vec{k},\vec{q})$ which describe the probability amplitude of an electron absorbing a phonon of mode $\nu$ and wave vector $\vec{q}$ and scattering from state $\ket{n,\vec{k}}$ to state $\ket{m,\vec{k}+\vec{q}}$ \cite{zhou2018electron, giustino2017electron}. A schematic diagram is given in Figure~\ref{fig:ife_and_vortex}(a). Following Ref. \cite{giustino2017electron}, the electron-phonon matrix elements are given by
\begin{multline}\label{matrix_elements}
        g_{mn\nu}(\vec{k}, \vec{q}) = \bra{m, \vec{k}+\vec{q}} \sum_{p} l_{\vec{q}\nu} e^{i\vec{q}\cdot\vec{R_p}}\frac{\partial U}{\partial u_{p\nu}}\ket{n,\vec{k}}.
\end{multline}
Using equation \eqref{displacement_mu_nu} and equation \eqref{matrix_elements}, we extend the single ion effective Hamiltonian \eqref{eq:H_eff_single_ion} and obtain the following effective Hamiltonian for the entire crystal,
\begin{multline}\label{eq:H_eff_crystal}
        \mathcal{H}_{\text{eff}}^{ab}(\vec{k}) = -i \hbar \omega\sum_{\vec{q}}\left[(\hat{a}_{\vec{q},\mu} + \hat{a}_{-\vec{q},\mu}^\dagger)(\hat{a}_{-\vec{q},\nu}^{\dagger}+\hat{a}_{\vec{q},\nu}) \vphantom{\frac{)g_{bn\nu}^*(\vec{k},\vec{q})}{\omega_a^2}}\right. \\ \left.
       \times \sum_n \frac{g_{an\mu}(\vec{k},\vec{q})g_{bn\nu}^*(\vec{k},\vec{q})-g_{an\nu}(\vec{k},\vec{q})g_{bn\mu}^*(\vec{k},\vec{q})}{E_{\vec{k}nb}^2-\hbar^2\omega^2} \right].
\end{multline}
We note that equation \eqref{eq:H_eff_crystal} makes no assumptions on the material and represents the main theoretical result of our paper.

Furthermore, we can show that (\ref{eq:H_eff_crystal}) can be connected to the phonon number operator. We introduce operators $\hat{a}_{\vec{q}}$, $\hat{a}_{\vec{-q}}^{\dagger}$, such that $\vec{\varepsilon}\hat{a}_{\vec{q}} = \left(\begin{smallmatrix} \hat{a}_{\vec{q},\mu} \\ \hat{a}_{\vec{q},\nu}\end{smallmatrix}\right)$, $\vec{\varepsilon^*}\hat{a}_{\vec{-q}}^{\dagger} = \left(\begin{smallmatrix} \hat{a}_{\vec{-q}, \mu}^{\dagger} \\ \hat{a}_{\vec{-q}, \nu}^{\dagger}\end{smallmatrix}\right)$. Now, using bosonic anticommutation relations, we can reformulate (\ref{eq:H_eff_crystal}) as
\begin{multline}\label{H_eff_number_operator}
    \mathcal{H}_{\text{eff}}^{ab}(\vec{k}) = -2i \hbar \omega\sum_{\vec{q}} \left[ \left(\hat{a}^{\dagger}_{-\vec{q}}\hat{a}_{\vec{q}}+\frac{1}{2}\delta_{-\vec{q},\vec{q}}\right)\right.  \\ \left.
       \times \sum_n \frac{g_{an\mu}(\vec{k},\vec{q})g_{bn\nu}^*(\vec{k},\vec{q})-g_{an\nu}(\vec{k},\vec{q})g_{bn\mu}^*(\vec{k},\vec{q})}{E_{\vec{k}nb}^2-\hbar^2\omega^2} \right],
\end{multline}
where, following Zhang and Niu \cite{Zhang2014}, we have omitted $\hat{a}_{\vec{q}}\hat{a}_{\vec{q}}$ and $\hat{a}^{\dagger}_{\vec{-q}}\hat{a}^{\dagger}_{\vec{-q}}$ terms. Additionally, we have used polarization vector property $\varepsilon^*_{\mu}\varepsilon_{\nu} = \delta_{\mu \nu}$, and the fact that for degenerate phonon modes that equally contribute to a circularly polarized mode we can set $\varepsilon_\mu = \varepsilon_\nu$. 
For a soft mode at $\Gamma$-point $\hat{a}^{\dagger}_{-\vec{q}}\hat{a}_{\vec{q}}$ becomes $\hat{n}_{\vec{0}} = \hat{a}^{\dagger}_{\vec{0}}\hat{a}_{\vec{0}}$, which is the occupation number operator for circularly polarized phonon mode at $\Gamma$-point.

\begin{figure}
    \centering
    \includegraphics[width=0.49\textwidth]{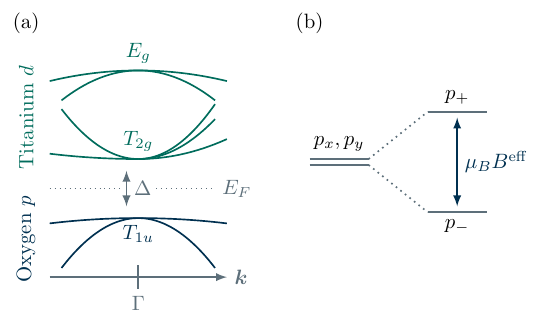}
    \caption{Schematic of the electronic structure of SrTiO$_3$. (a) Orbital character of valence and conduction bands. (b) Splitting of $p$-orbitals in the effective magnetic field. }
    \label{fig:elstruc}
\end{figure}

Relating to the recent finding of the large dynamical multiferroicity in SrTiO$_3$~\cite{Basini2024} we discuss the effective Hamiltonian \eqref{eq:H_eff_crystal} for cubic symmetry and an infrared active optical phonon mode with $T_{1u}$ symmetry. The electronic structure of SrTiO$_3$ is schematically shown in Figure~\ref{fig:elstruc}. The valence band is primarily composed of oxygen $p$-states ($T_{1u}$) and the conduction band is composed of Ti-$d$ states ($T_{2g}$). To estimate the effective magnetic field imposed on the electrons by axial phonons, we discuss the level splitting of $\Gamma$-point states transforming as $p_\pm = (p_x \pm i p_y)/\sqrt{2}$. Hence, we evaluate the overlap elements $\mathcal{H}_{\text{eff}}^{xy}(\vec{0})$ and $\mathcal{H}_{\text{eff}}^{yx}(\vec{0})$ in the effective Hamiltonian \eqref{eq:H_eff_crystal}. Here, we use two assumptions. First, the electron-phonon coupling elements $g_{an\nu}$ are subject to selection rules~\cite{Shu2024,Chen2021}. As both, the phonon mode and the $p$-orbitals are parity-odd, the overlap needs to involve parity-even orbitals, i.e., the Ti-$d$-states. Second the perturbative sum in \eqref{eq:H_eff_crystal} rapidly decreases with the spectral distance, $\mathcal{H}_{\text{eff}}^{ab} \sim E_{nb}^{-2}$ for $E_{nb}\gg \hbar\omega$. Using these assumptions and cubic symmetry, we derive (details given in the supplementary materials),
\begin{align}
    \mathcal{H}_{\text{eff}}^{xy}(\vec{0}) &= - \mathcal{H}_{\text{eff}}^{yx}(\vec{0}), \\
    \mathcal{H}_{\text{eff}}^{xy}(\vec{0}) &=  -i\left(\hat{n}_{\vec{0}}+\frac{1}{2}\right) \frac{\hbar \omega \abs{g}^2}{\Delta^2-\hbar^2\omega^2}. \label{H_eff_gamma}
\end{align}
A basis transform to $p_\pm = \left(p_x \pm i p_y\right)/\sqrt{2}$, i.e., $H_{\text{eff}}^{\pm\pm} = \pm i \mathcal{H}_{\text{eff}}^{xy}$ gives a two-phonon amplitude
\begin{equation}
    E^\pm = \pm \frac{\hbar \omega \abs{g}^2}{\Delta^2-\hbar^2\omega^2}\left(n_{\vec{0}}+\frac{1}{2}\right). 
\end{equation}
At room temperature, the infrared-active ferroelectric soft-mode in SrTiO$_3$ has a frequency of 2.7~THz~\cite{Basini2024,Voigt1995SrTiO3KTaO3}, i.e., $\hbar \omega \approx 11~\text{meV}$. In contrast, the measured direct band gap of SrTiO$_3$ is $\Delta \approx 3.75~\text{eV}$~\cite{bentham2001}. If we assume an electron phonon coupling of $g \approx 7~\text{meV}$~\cite{Zhou2018, Gastiasoro2023}, we obtain $\Delta E = E^+ - E^- \approx 7.7\times 10^{-8}\left(n_{\vec{0}}+\frac{1}{2}\right)~\text{eV}$. If we compare this to the expected Zeeman splitting due to a magnetic field, 
\begin{equation}\label{splitting}
    \frac{\Delta E}{2} = g_J \mu_B B_z^{\text{eff}},
\end{equation}
we can estimate the effective magnetic field for $J=1$. Relating to the measurements done on SrTiO$_3$ \cite{Basini2024}, we first estimate the number of phonons resulting from the pump field strength of $E_0 = 230$~kV/cm that was used in the experiment. The intensity of laser field is given by $I = \frac{1}{2\mu_0 c}E_0^2$. With the energy of a single photon, we can estimate the number of incident photons per area of a unit cell as $N \approx 60$~1/ps. Given a pulse width of 2~ps, that results in the total number of incident of photons per unit cell area $N^{\text{ph}} = 119$. At the same time, at the resonance frequency we can assume that one photon is able to excite one phonon. Therefore $N^{\text{ph}}\approx120$ will excite approximately 60 pairs of phonons at the surface, i.e. $n_{\vec{0}}=60$. Thus, from \eqref{splitting} we can estimate the effective magnetic field to be $B^{\text{eff}} \approx 40~\text{mT}$. Basini \textit{et al.} \cite{Basini2024} report an effective magnetic field of 32~mT at the surface of the sample, therefore we can conclude that our estimate is in agreement with the experimental observations. Here we also want to highlight that since the number of phonons is proportional to the laser intensity, $n_{\vec{0}} \propto I$, the magnitude of the effect scales with the square of the pump's electric field strength, $E_0^2$.

We stress that the effective magnetic field is not a ''physical`` magnetic field as described by Maxwell's equations. Still, it provides a time-reversal symmetry breaking field. Recently, Merlin \cite{Merlin2023} pointed out that by probing a material with the magneto-optical Kerr effect, such a time-reversal symmetry breaking field, or non-Maxwellian field, leads to a Kerr rotation of the linearly polarized probe laser, with electric field $\vec{\varepsilon}$. This process can be described phenomenologically by the following free energy
\begin{equation}
    F_{\text{MOKE}} = i \Lambda B^{\text{eff}} \left(\vec{\varepsilon}\times \vec{\varepsilon}^*\right),
\end{equation}
where the Kerr rotation results from the difference of the dielectric constants for left- and right-circularly polarized light, in the presence of circularly polarzed phonons and a resulting effective magnetic field,
\begin{equation}
    \Delta\epsilon_R = -4\pi \Lambda B^{\text{eff}}, \qquad\Delta\epsilon_L = 4\pi \Lambda B^{\text{eff}}.
    \label{eq:FaradayRot}
\end{equation}
Equation \eqref{eq:FaradayRot} follows a similar derivation as shown in equation \eqref{eq:phononepsilon}~\cite{pershan1966theoretical}.

In the experiment reported by Basini \textit{et al.} \cite{Basini2024}, circularly polarized phonons in SrTiO$_3$ were induced by a circularly polarized laser field. Due screening in the material, the penetration depth of the laser pump-pulse, measured by the decay length $l_{\text{decay}}$ is in the order of $l_{\text{decay}} \approx 2.5~\mu m$. 
Hence, the expected Faraday rotation can be estimated as follows \cite{Basini2024, Freiser1968},
\begin{align}\label{Faraday_rotation}
    \theta_F &= \frac{l_{\text{decay}}V}{2} B^{\text{eff}}_z \\
    &= \frac{l_{\text{decay}}V}{2} \frac{1}{\mu_B} \frac{\hbar \omega \abs{g}^2}{\Delta^2-\hbar^2\omega^2}\left(n_{\vec{0}}+\frac{1}{2}\right).
\end{align}
Here, V is the Verdet constant of the material, which is $V \approx 180~\text{rad}\,\text{m}^{-1}\text{T}^{-1}$ for SrTiO$_3$ \cite{Basini2024}.

\begin{figure}
    \centering
    \includegraphics[width=0.4\textwidth]{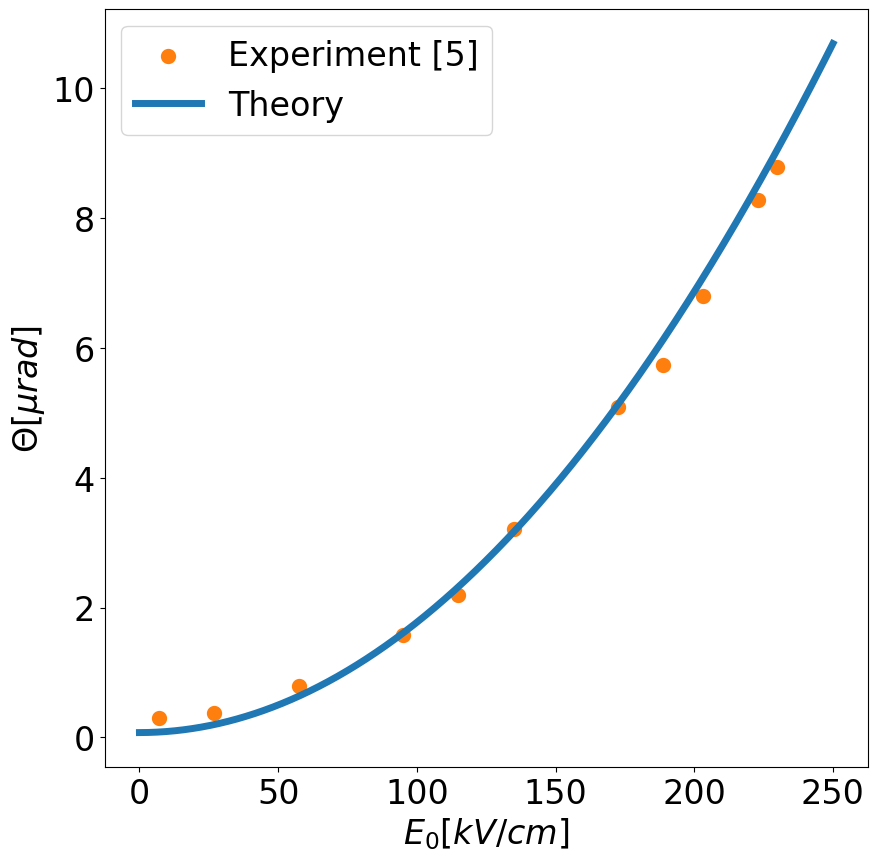}
    \caption{Polarization rotation as a function of electric field strength. Solid line is the Faraday rotation obtained with \eqref{Faraday_rotation} and $|g|=7\text{~meV}$. Dots represent Kerr rotation measurements reported by Basini \textit{et al.} \cite{Basini2024}. }
    \label{fig:Kerr rotation}
\end{figure}

Since Kerr and Faraday rotation are closely related and are not different by more than a factor of 2 in SrTiO$_3$ \cite{Basini2024}, an estimate of Faraday rotation gives a good picture approximation of Kerr rotation resulting from the same magnetic field. In Fig. (\ref{fig:Kerr rotation}) we plot Faraday rotation calculated using \eqref{Faraday_rotation} as a function of electric field strength $E_0$. Comparison to experimental measurements of Kerr rotation \cite{Basini2024} shows that for $|g| = 7\text{~meV}$ our estimate is in perfect agreement with the experimental values. Here we would like to emphasize again that our plot shows a quadratic dependency of the effect on the electric field strength.

\section{Discussion \& Summary}
To highlight the fact that phonon inverse Faraday effect arises from circularly polarized phonons, we consider the ladder operators $(\hat{a}_{\vec{0},\mu} + \hat{a}_{\vec{0},\mu}^\dagger)(\hat{a}_{\vec{0},\nu}^{\dagger}+\hat{a}_{\vec{0},\nu})$ in the effective Hamiltonian \eqref{H_eff_gamma}, and define $\hat{O}_\mu = \hat{a}_{\vec{0},\mu} + \hat{a}_{\vec{0},\mu}^\dagger$ and $\hat{O}_\nu = \hat{a}_{\vec{0},\nu} + \hat{a}_{\vec{0},\nu}^\dagger$, respectively. These operators describe the amplitude of the corresponding phonon modes $\mu$ and $\nu$. Similarly to Ref. \cite{chaudhary2023giant}, we introduce operators $\hat{O}_{\pm} = \left(\hat{O}_{\mu} \pm i \hat{O}_{\nu}\right)/\sqrt{2}$, which allows us to write the effective Hamiltonian \eqref{H_eff_gamma} as
\begin{equation}
    \mathcal{H}_{\text{eff}}^{\pm\pm}(\vec{0}) = \mp\frac{i}{4} \left(\hat{O}_+\hat{O}_+-\hat{O}_-\hat{O}_-\right) \frac{\hbar \omega \abs{g}^2}{\Delta^2-\hbar^2\omega^2}.
\end{equation}

Finally, we would like to relate our result to other theories on the problem. We note that equation \eqref{eq:H_eff_crystal} is a generalization of the orbit-lattice coupling described by Chaudhary \textit{et al.} \cite{chaudhary2023giant}. In the 4$f$ paramagnets such as CeCl$_3$, the spectral distance $E_{nb}$ is dominated by either spin-orbit interaction ($\approx 0.1$~eV) or by the crystal field splitting ($6~\text{meV}$). As a result, the expected effective magnetic field is significanty larger as compared to SrTiO$_3$. In fact, this is consistent with experimental work \cite{Schaack1977,Luo2023} as well as theoretical estimates~\cite{chaudhary2023giant,Juraschek2022}. The argument of tiny spectral gaps due to crystal field effects is also found in connection to a dynamical crystal field effect imposed by the phonon \cite{Klebl2024}. Furthermore, for materials with large gap $\Delta \gg \hbar \omega$, the denominator of the effective magnetic field becomes independent of the phonon frequency, i.e., $\Delta^2 - \hbar^2\omega^2 \approx \Delta^2$. In this limit, the level splitting is linearly dependent of the phonon frequency, $\Delta E \sim \hbar \omega$, which is related to the inertial effects discussed in Refs. \cite{geilhufe2022dynamic,geilhufe2023electron}. The same strong suppression of the expected effective magnetic field or magnetization by the band gap, $B^{\text{eff}} \sim \Delta^{-2}$, found in the present work, is also revealed in the formalism of the modern theory of magnetization and the phonon magnetic moment from  electronic topology \cite{ren2021phonon,Saparov2022}. In particular, in the adiabatic regime, equation \eqref{eq:H_eff_crystal} becomes identical to the formalism developed by Ren \textit{et al.} \cite{ren2021phonon}.

In summary, the approach presented here provides a general and material-independent framework for estimating an emergent magnetization and effective magnetic field due to axial phonons, i.e., phonons carrying angular momentum~\cite{Zhang2014}. Furthermore, the result given by equation \eqref{eq:H_eff_crystal} highlights an important distinction between the phonon inverse Faraday effect and the optical inverse Faraday effect. While the microscopic theory of the optical inverse Faraday is based on the dipole coupling of the electric field to the electron, the phonon inverse Faraday effect is based on the electron-phonon interaction. As such the phonon inverse Faraday effect also occurs in the absence of a laser field, as long as an imbalance of left- and right-circularly polarized phonons is present. However, it is also worth noting that the circularly polarized phonons can be induced by a circularly polarized laser field \cite{Basini2024}. Hence, for laser excitations resonant with phonons, both the phononic and the optical contribution coexist. 

\begin{acknowledgments}
We acknowledge inspiring discussions with Dominik Juraschek, Hanyu Zhu, Martina Basini, Stefano Bonetti, Alexander Balatsky, Finja Tietjen. 
Also, We acknowledge support from the Swedish Research Council (VR starting Grant No. 2022-03350), the Olle Engkvist Foundation, the Royal Physiographic Society in Lund and Chalmers University of Technology via the Department of Physics, and the areas of advance Nano and Materials. 

In the process of finalizing this work, another interesting paper appeared discussing a similar approach \cite{Merlin2024}.
\end{acknowledgments}

\clearpage
\newpage
\widetext

\begin{center}
\textbf{\large Supplementary Materials: Phonon Inverse Faraday effect from electron-phonon coupling}\\[1.2ex]
Natalia Shabala and R. Matthias Geilhufe\\[1.2ex]
\textit{Department of Physics, Chalmers University of Technology, 412 96 G\"{o}teborg, Sweden}
\end{center}

\section*{Single ion}

The perturbation due to lattice vibrations has the form
\begin{equation}
    V(t) = 2\Re\left[\vec{u}\cdot\nabla_{\vec{u}}U\right].
\end{equation}
It can be written as harmonic perturbation $V(t) = v e^{i\omega t} + v^*e^{-i\omega t}$, with perturbation amplitude $v$ given by 
\begin{equation}
    v = \vec{u} \cdot \nabla U.
\end{equation}
We define two potential operators $U_{\pm}$, as $U_\pm = \left(\frac{\partial U}{\partial u_x} \pm i \frac{\partial U}{\partial u_y}\right)/\sqrt{2}$. Thus the perturbation amplitude can be expressed as
\begin{equation}\label{pert_amp_single_ion}
    v = u_R\nabla U_+ + u_L \nabla U_{-}.
\end{equation}
Now we can derive the expression for the effective Hamiltonian starting with the equation derived by Pershan et al. \cite{pershan1966theoretical}
\begin{equation}\label{H_eff_Pershan}
    \bra{a}H_{\text{eff}}(t)\ket{b} = -\sum_n \left[\frac{\bra{a}v\ket{n}\bra{n}v^*\ket{b}}{E_{nb}-\hbar\omega} - \frac{\bra{a}v^*\ket{n}\bra{n}v\ket{b}}{E_{nb}+\hbar\omega}\right].
\end{equation}
The form of the perturbation amplitude in equation (\ref{pert_amp_single_ion}) combined with the effective Hamiltonian (\ref{H_eff_Pershan}), gives us
\begin{equation}
\begin{split}
    \mathcal{H}_{\text{eff}}^{ab} =   - &  \sum_n \Big[\frac{E_{nb}}{E_{nb}^2-\hbar\omega^2}\Big((u_Ru_R^*+u_Lu_L^*)(\nabla U_{+}^{an} \nabla U_{-}^{an} + \nabla U_{-}^{an} \nabla U_{+}^{an}) + \\
    + & 2u_Lu_R^*\nabla U_{-}^{an}\nabla U_{-}^{an} + 2u_L^*u_R\nabla U_{+}^{an}\nabla U_{+}^{an}\Big) + \\
    + & \frac{\hbar\omega}{E_{nb}^2 - \hbar^2\omega^2}(u_Ru_R^*-u_Lu_L^*)(\nabla U_{+}^{an} \nabla U_{-}^{an} - \nabla U_{-}^{an} \nabla U_{+}^{an})\Big].
\end{split}
\end{equation}
Here, only $(u_Ru_R^*-u_Lu_L^*)$ transforms as a magnetic field therefore this term is the only relevant term and all others can be discarded. In Cartesian coordinates this gives us
\begin{equation}
     \mathcal{H}_{\text{eff}}^{ab} =  -(\vec{u} \times \vec{u}^{*})_z \sum_n \frac{\hbar\omega}{E_{nb}^2-\hbar^2\omega^2} ( \nabla U^{an} \times \nabla U^{nb} )_z.
\end{equation}

\section*{Entire crystal}

To derive the expression for the effective Hamiltonian for the entire crystal, we introduce phonon displacement amplitudes $u_{p \mu}$ and $u_{p' \nu }$ for phonon modes $\mu$, $\nu$ and unit cells $p$ and $p'$. Then we can use these phonon modes to define the effective Hamiltonian by analogy with \eqref{eq:H_eff_single_ion}:
\begin{equation}\label{H_eff_crystal(1)}
    \mathcal{H}_{\text{eff}}^{ab} = -\frac{\hbar\omega}{4}\sum_{n,p,p'} (u_{p\mu}u_{p'\nu}^*-u_{p'\nu}u_{p\mu}^*)\frac{\bra{a}\frac{\partial U}{\partial u_{p\mu}}\ket{n}\bra{n}\frac{\partial U}{\partial u_{p'\nu}}\ket{b}-\bra{a}\frac{\partial U}{\partial u_{p'\nu}}\ket{n}\bra{n}\frac{\partial U}{\partial u_{p\mu}}\ket{b}}{E_{nb}^2-\hbar^2\omega^2}.
\end{equation}
Given \eqref{circularphonon}, for circularly polarized phonons we quantize the phonon displacement in the following way
\begin{equation}
\begin{split}
    & u_{p \mu} =  i\sum_{\vec{q}}e^{i\vec{q}\cdot \vec{R_p}}l_{\vec{q}\mu}(\hat{a}_{\vec{q}\mu}+\hat{a}^{\dagger}_{-\vec{q}\mu}) \\
    & u_{p \nu} =  \sum_{\vec{q}}e^{i\vec{q}\cdot \vec{R_p}}l_{\vec{q}\nu}(\hat{a}_{\vec{q}\nu}+\hat{a}^{\dagger}_{-\vec{q}\nu})
\end{split}
\end{equation}
where $l_{\vec{q}\nu}=\sqrt{\frac{\hbar}{2\omega_{\vec{q}\nu}}}$ denotes the zero displacement amplitude. Operators $\hat{a}_{\vec{q}\nu}^{\dagger}$ and $\hat{a}_{\vec{q}\nu}$ are bosonic creation and annihilation operators and operator $(\hat{a}_{\vec{q}\nu}+\hat{a}^{\dagger}_{-\vec{q}\nu})$ can be interpreted as phonon displacement operator. 

Here it is useful to introduce electron-phonon matrix elements $g_{mn\nu}(\vec{k},\vec{q})$ \cite{giustino2017electron}:
\begin{equation}
        g_{mn\nu}(\vec{k}, \vec{q}) = \bra{m, \vec{k}+\vec{q}} \sum_{p} l_{\vec{q}\nu} e^{i\vec{q}\cdot\vec{R_p}}\frac{\partial U}{\partial u_{p\nu}}\ket{n,\vec{k}}.
\end{equation}
With equation \eqref{displacement_mu_nu} and equation \eqref{matrix_elements} we can express the effective Hamiltonian \eqref{H_eff_crystal(1)} in terms of electron-phonon matrix elements $g_{an\nu}(\vec{k}, \vec{q})$, $g_{bn\mu}(\vec{k}, \vec{q'})$ and quantized displacement operators $(\hat{a}_{\vec{q}\mu}+\hat{a}^{\dagger}_{-\vec{q}\mu})$, $(\hat{a}_{\vec{q'}\nu}+\hat{a}^{\dagger}_{-\vec{q'}\nu})$. We consider an effective Hamiltonian that is diagonal in wave vector $\vec{k}$. Therefore in $\bra{a, \vec{k}}e^{i\vec{q}\cdot\vec{R_p}}\frac{\partial U}{\partial u_{p\mu}}\ket{n, \vec{k'}}\bra{n, \vec{k'}}e^{-i\vec{q'}\cdot\vec{R_{p'}}}\frac{\partial U}{\partial u_{p'\nu}}\ket{b, \vec{k}}$ it must be true that $\vec{k}=\vec{k'}+\vec{q}$ and $\vec{k'}-\vec{q'}$. Thus $\vec{q}=\vec{q'}$. Thus the effective Hamiltonian is given by
\begin{equation}
        \mathcal{H}_{\text{eff}}^{ab}(\vec{k}) = -i \frac{\hbar \omega}{2}\sum_{\vec{q}}\left[(\hat{a}_{\vec{q},\mu} + \hat{a}_{-\vec{q},\mu}^\dagger)(\hat{a}_{-\vec{q},\nu}^{\dagger}+\hat{a}_{\vec{q},\nu}) \vphantom{\frac{)g_{bn\nu}^*(\vec{k},\vec{q})}{\omega_a^2}}\right. \\ \left.
       \times \sum_n \frac{g_{an\mu}(\vec{k},\vec{q})g_{bn\nu}^*(\vec{k},\vec{q})-g_{an\nu}(\vec{k},\vec{q})g_{bn\mu}^*(\vec{k},\vec{q})}{E_{\vec{k}nb}^2-\hbar^2\omega^2} \right].
\end{equation}

\section*{Effective magnetic field}
We consider cubic symmetry and the interaction of electronic states with a $\Gamma$-point infrared-active phonon, transforming as the irreducible representation $T_{1u}$. The electron-phonon coupling elements are subject to selection rules\cite{Chen2021,Shu2024}, where electrons from an orbital transforming as $D^n$ are allowed to scatter to an orbital transforming as $D^a$ iff $D^a$ occurs on the right-hand side of the Clebsch-Gordan decomposition, 
\begin{table}[b]
    \centering
    \begin{tabular}{ll|ccc|ccc|ccc}
    \hline\hline
    electron & $T_{1u}$ & $p_x$ &  & & $p_y$ &  &  & $p_z$ &  &  \\
    phonon & $T_{1u}$ & x & y & z & x & y & z & x & y & z \\
    \hline
      $T_{2g}$ & xz & 0 & 0 & $\frac{1}{\sqrt{2}}$ & 0 & 0 & 0 & $\frac{1}{\sqrt{2}}$ & 0 & 0 \\
     & yz & 0 & 0 & 0 & 0 & 0 & $\frac{1}{\sqrt{2}}$ & 0 & $\frac{1}{\sqrt{2}}$ & 0 \\
      & xy & 0 & $\frac{1}{\sqrt{2}}$ & 0  & $\frac{1}{\sqrt{2}}$ & 0 & 0 & 0 & 0 & 0 \\
    \hline\hline
    \end{tabular}
    \caption{Relevant Clebsch-Gordan coefficients}
    \label{tab:cgc}
\end{table}
\begin{equation}
    T_{1u} \otimes D^n \simeq N_1 D^1 \oplus N_2 D^2 \oplus \dots \label{suppl:cgs}
\end{equation}
$N_a$ is the number of times $D^a$ occurs within the Clebsch-Gordan sum \eqref{suppl:cgs}. The electron-phonon matrix element relates to the Wigner-Eckart theorem~\cite{gtpack2},
\begin{equation}
    g_{an\nu} = g_{D^a \alpha; D^n \eta; T_{1u} \nu} = \sum_{\xi = 1}^{N_a} \left(\overset{D^n}{\eta} \overset{T_{1u}}{\nu} \right|\left.\overset{D^a,}{\alpha}\,\overset{\xi}{\vphantom{\nu}} \right)^* \left(D^a|T_{1u}|D^n\right),
\end{equation}
with the sum running over a product of Clebsch-Gordan coefficients and reduced matrix elements. We estimate the effective magnetic field in SrTiO$_3$ by assuming oxygen $p$-states. The $p$ states at the $\Gamma$-point ($\vec{k}=0$) transform as the irreducible representation $T_{1u}$. We decompose the direct product of the irreducible representations of the $p$-orbitals ($T_{1u}$) and the infrared active phonon mode ($T_{1u}$), 
\begin{equation}
    T_{1u}\otimes T_{1u} \sim A_{1g} \oplus T_{1g} \oplus T_{2g} \oplus E_g. 
\end{equation}
Hence, the overlap is only non-zero if the $p$-electron scatters into an $s$ or $d$ state, with representations $A_{1g}$ or $E_g$ and $T_{2g}$, respectively. Furthermore, we notice that $N_n = 1$ in either case. According to equation \eqref{eq:H_eff_crystal}, the perturbative series is suppressed by the size of the spectral distance squared, $~1/\omega_{nb}^2$, $\omega \ll \omega_{nb}$ and it suffices to only consider the overlap between $T_{1u}$ and $T_{2g}$ states, corresponding to the band gap $\Delta \approx 3.25~\text{eV}$~\cite{bentham2001}. Hence, the overlap element for $p_x$ and $p_y$ in the effective Hamiltonian \eqref{eq:H_eff_crystal}, for electrons at $\vec{k}=0$ and phonons at $\vec{q}=0$, is given by
\begin{multline}
        \mathcal{H}_{\text{eff}}^{xy}(\vec{k}=0) = -i(\hat{a}_{\vec{0},\nu} + \hat{a}_{\vec{0},\nu}^\dagger)(\hat{a}_{\vec{0},\mu}^{\dagger}+\hat{a}_{\vec{0},\mu}) \frac{\hbar \omega \abs{g}^2}{\Delta^2-\hbar^2\omega^2}\sum_\eta \left[
\left(\overset{T_{2g}}{\eta} \overset{T_{1u}}{\nu} \right|\left.\overset{T_{1u}}{x}\right)^*\left(\overset{T_{2g}}{\eta} \overset{T_{1u}}{\mu} \right|\left.\overset{T_{1u}}{y}\right)-
\left(\overset{T_{2g}}{\eta} \overset{T_{1u}}{\mu} \right|\left.\overset{T_{1u}}{x}\right)^*\left(\overset{T_{2g}}{\eta} \overset{T_{1u}}{\nu} \right|\left.\overset{T_{1u}}{y}\right) \right].\label{Hamiltonian-xy}
\end{multline}
Here we introduced $g$ as the reduced matrix element. Equation \eqref{Hamiltonian-xy} can be evaluated from the Clebsch-Gordan coefficients calculated using GTPack~\cite{gtpack2,gtpack1} and given in Table~\ref{tab:cgc},
\begin{equation}
    \mathcal{H}_{\text{eff}}^{xy}(\vec{k}=0) = - \mathcal{H}_{\text{eff}}^{yx}(\vec{k}=0) = -\frac{i}{2}(\hat{a}_{\vec{0},\nu} + \hat{a}_{\vec{0},\nu}^\dagger)(\hat{a}_{\vec{0},\mu}^{\dagger}+\hat{a}_{\vec{0},\mu}) \frac{\hbar \omega \abs{g}^2}{\Delta^2-\hbar^2\omega^2}.
\end{equation}
Alternatively the equation above can be written in a form similar to (\eqref{H_eff_number_operator}):
\begin{equation}
    \mathcal{H}_{\text{eff}}^{xy}(\vec{k}=0) = - \mathcal{H}_{\text{eff}}^{yx}(\vec{k}=0) = -i(\hat{n}_{\vec{0}}+\frac{1}{2}) \frac{\hbar \omega \abs{g}^2}{\Delta^2-\hbar^2\omega^2},
\end{equation}
where $\hat{n}_{\vec{0}}$ is phonon number operator.

\section*{Phonon normal coordinates}

Phonon displacement for an atom $j$ in a unit cell $p$ in Cartesian direction $\alpha$ can be quantized and written in terms of bosonic creation and annihilation operators $\hat{a}_{\vec{q}\nu}$, $\hat{a}^{\dagger}_{\vec{\nu}\vec{q}}$. The expression then takes the form \cite{giustino2017electron}
\begin{equation}
    \Delta \tau_{p j \alpha} = \left(\frac{m_0}{N_p m_j} \right)^{\frac{1}{2}} \sum_{\vec{q}\nu}e^{i\vec{q}\cdot\vec{R_p}}\xi_{j\alpha p}(\vec{q})\Tilde{l}_{\vec{q}\nu}(\hat{a}_{\vec{q}\nu}+\hat{a}_{-\vec{q}\nu}^{\dagger}).
\end{equation}
Here, $N_p$ is the total number of unit cells, $m_j$ - mass of the ion $j$, $m_0$ - an arbitrary reference mass, typically taken to be equal to the proton mass. We also use the phonon polarization vector $\xi_{j\alpha p}$ and $\Tilde{l}_{\vec{q}\nu}$ - zero displacement amplitude given by $\sqrt{\frac{\hbar}{2m_0\omega_{\vec{q}\nu}}}$. 

We can transform this expression to obtain phonon normal coordinates given by $u_{p\nu}=\sum_{j\alpha}m_j \xi_{j\alpha \nu}(\vec{q}) \Delta \tau_{p j \alpha}$ \cite{Basini2024}. Thus for a unit cell $p$ we obtain
\begin{equation}
    u_{p\nu} = \sum_{j\alpha} \sqrt{m_0} \xi_{j\alpha \nu}(\vec{q}) \sum_{\vec{q}\nu}e^{i\vec{q}\cdot\vec{R_p}}\xi_{j\alpha \nu'}(\vec{q})l_{\vec{q}\nu}(\hat{a}_{\vec{q}\nu}+\hat{a}_{-\vec{q}\nu}^{\dagger}).
\end{equation}
For real polarization vectors $\sum_{j\alpha}\xi_{j\alpha\nu}(\vec{q})\xi_{j\alpha\nu'}(\vec{q})=\delta_{\nu\nu'}$. We can thus write
\begin{equation}
    u_{p\nu} = \sum_{\vec{q}} e^{i\vec{q}\cdot\vec{R_p}}\sqrt{\frac{\hbar}{2\omega_{\vec{q}\nu}}}(\hat{a}_{\vec{q}\nu}+\hat{a}_{-\vec{q}\nu}^{\dagger}).
\end{equation}
At $\Gamma$-point this will give us 
\begin{equation}
    u_{p\nu} = l_{\vec{q}\nu}(\hat{a}_{\vec{0}\nu}+\hat{a}_{\vec{0}\nu}^{\dagger}),
\end{equation}
where $l_{\vec{0}\nu}=\sqrt{\frac{\hbar}{2\omega_{\vec{0}\nu}}}$

\end{document}